\documentclass[twocolumn,notitlepage,floats,nofootinbib,amsmath,amssymb,aps,prd]{revtex4-1}

\usepackage{amsmath,amssymb,amsfonts,latexsym,cancel,amsthm}

\usepackage{mathtools}
\usepackage{graphicx}
\usepackage{color}
\usepackage{amsbsy}
\usepackage{hyperref}
\usepackage{tikz}
\usepackage{comment}
\usepackage{seqsplit}

\newcommand{\be}{\begin{equation}}
\newcommand{\beq}{\begin{equation}}
\newcommand{\ee}{\end{equation}}

\def\bea {\begin{eqnarray}}
\def\eea {\end{eqnarray}}

\def\dd{{\rm d}}

\definecolor{newgreen}{rgb}{0.0, 0.75, 0.0}
\definecolor{cadmiumgreen}{rgb}{0.0, 0.42, 0.24}

\begin{document}

\title{On weak solutions in Einstein theory and beyond }

\author{Francesco Fazzini} \email{francesco.fazzini@unb.ca}
\affiliation{Department of Mathematics and Statistics, University of New Brunswick, \\
Fredericton, NB, Canada E3B 5A3}

\author{Hassan Mehmood} \email{hassan.mehmood@unb.ca}
\affiliation{Department of Mathematics and Statistics, University of New Brunswick, \\
Fredericton, NB, Canada E3B 5A3}

\begin{abstract}
In spherical symmetry, gravitational collapse of dust may give rise to the so-called shell-crossing singularities, beyond which spacetime can be extended using weak solutions to the integrated version of the equations of motion. We argue that the paradigm of weak solutions is ill-suited for dynamical extension beyond shell crossings through shock waves,
because it is based implicitly on the assumption that the Painlevé-Gullstrand time remains continuous across the shock, and leads to the unwanted prospect of shock waves of dust particles moving faster than light. 
\end{abstract}

\maketitle

\section{Introduction}
The problem of gravitational collapse of matter has drawn significant interest since the inception of general relativity. One of the earliest attempts at understanding the problem was the pioneering work of Oppenheimer and Snyder \cite{Oppenheimer:1939ue}, in which they showed that a spherically symmetric and homogeneous distribution of pressureless fluid (i.e. dust) continues to collapse until it forms a black hole with a singularity at the center. Since then, this picture of gravitational collapse as an ill-fated journey unto a singularity has been confirmed from a number of perspectives, ranging from the time-honored singularity theorems to numerical studies of matter collapse using different types of matter. Within the context of dust collapse, the Oppenheimer-Snyder model can be generalized to inhomogeneous distributions of dust in the framework of Lemaitr\'e-Tolman-Bondi (LTB) spacetimes \cite{Bondi:1947fta, Tolman:1934za}.

It is widely believed that quantum gravity can resolve the singularities that are endemic to general relativity. However, what should replace the singularity in a gravitational collapse scenario is a matter of open debate. One interesting possibility is that the singularity is replaced by a bounce: matter falls inward, crosses an apparent horizon, thus forming a black hole, and then reaches a minimal radius, at which point it starts expanding and eventually exits the apparent horizon. The idea is inspired by loop quantum gravity (LQG), which predicts the geometry of spacetime to be discrete at a microscopic level and hence posits the existence of a minimal length. It has been concretely realized in phenomenological or effective models of gravitational collapse \cite{Husain:2022gwp,Bobula:2023kbo,Lewandowski:2022zce,Husain:2021ojz, Cipriani:2024nhx, Giesel:2023hys,Alonso-Bardaji:2023qgu,Han:2023wxg,Cafaro:2024vrw} which incorporate LQG-inspired quantum corrections to classical models of inhomogeneous dust collapse in the framework of LTB spacetimes. 

One curious feature of collapsing dust in LTB spacetimes is the existence of the so-called shell-crossing singularities (SCS) \cite{chao-hao, hu, yodzis1973occurrence}. These are points where spherical shells of matter collide with each other. At such points, the energy density of matter diverges, and the Einstein equations break down. 
 A way out of this problem, both in the classical theory \cite{Nolan:2003wp, Lasky:2006hq} and in the effective theory \cite{Husain:2022gwp, Cipriani:2024nhx}, is to study the problem in generalized Painlev\'e-Gullstrand (PG) coordinates \cite{Martel:2000rn}. In these coordinates, the equations of motion turn out to be hyperbolic balance laws and SCS are nothing but crossings of characteristics for these equations. Thus, familiar techniques from the theory of hyperbolic balance laws \cite{leveque-92, Dafermos:1315649} can be used to extend the spacetime beyond SCS. These methods consist in finding weak solutions to the equations of motion, i.e., solutions to the integrated version of the equations. As a result, the SCS are replaced by shock waves, which are propagating discontinuities in the variables entering the equations of motion.

While this is a quite reasonable approach to dynamical extension beyond SCS, it suffers from a rather serious ambiguity. As pointed out by its progenitor \cite{Nolan:2003wp} in the classical theory, and by \cite{Fazzini:2023ova} in the effective theory, weak solutions are not unique. In particular, different ways of writing down the integrated equations of motion may lead to distinct weak solutions to the same initial data, despite the fact that smooth solutions to those equations are identical. This is a familiar phenomenon in the theory of hyperbolic conservation laws \cite{leveque-92, Dafermos:1315649}, and is connected with the mathematical impossibility of defining a product of distributions \cite{colombeau-92}. In typical applications of the theory, such as in fluid dynamics, this type of non-uniqueness is not a problem\footnote{There is another type of non-uniqueness that is endemic to hyperbolic conservation laws as well: it concerns the existence of distinct weak solutions given the same initial data for \textit{one particular conservation equation written in terms of one particular set of variables}. Finding the unique solution in this case has constituted an area of deep and intense mathematical research in the twentieth century, bringing together insights from physics, functional analysis, distribution theory, and so on. For a detailed exposition of the subject, see \cite{Dafermos:1315649}.}, since the equations of motion are formulated as physically motivated integral equations to begin with; thus, changing the variables and getting different integral equations would be ad hoc. However, in gravitational dust collapse, the equations of motion are fundamentally differential equations, and different ways of arriving at them result in their differing up to multiplicative factors \cite{Fazzini:2025hsf}. There is no a priori justification for picking one of these differential equations over others for the purpose of integrating them to find weak solutions. 

As a way out of this problem, one might recall the adage ``gravity is geometry'' and attempt to find a geometrical criterion to pick one weak solution out of a zoo of them. As pointed out above, weak solutions consist of propagating discontinuities in the variables that enter the equations of motion. This suggests that there is a discontinuity in the metric as well. However, the metric contains these variables in a particular combination, and it is not necessary that a combination of discontinuous functions be discontinuous too. Thus, one can insist that only those formulations of the integrated equations of motion are to be favored which lead to the metric being continuous. This is the proposal put forth in \cite{Husain:2025wrh, Liu:2025fil}, and leads to a unique weak solution. However, this comes at a price: we shall show that a continuous weak solution to the equations of dust collapse in spherical symmetry gives rise to a shock wave that can move faster than light. In the classical theory, the shock, whatever its speed, has to eventually countenance oblivion in the central singularity of the black hole. However, in the effective theory, crucial details of a star's life, such as its lifetime, its post-bounce fate (whether it will form a regular black hole remnant, undergo a black-hole-to-white-hole transition, etc.) depend sensitively on the speed of the shock wave. This raises doubts as to the viability of weak solutions.

In addition to pointing out the unwanted feature of superluminal shocks in weak solutions, we will also explore the reason for it. The problem, we shall find out, is intimately connected with the fact that one is attempting to use a single PG chart to describe the entire spacetime. In particular, we shall show that implicit in the weak formulation is the assumption that the PG time coordinate stays continuous across the shock as a function of the proper time of an observer comoving with the shock. This suggests that one cannot perhaps use PG coordinates to describe the entire spacetime during dust collapse, a view also advanced in \cite{Fazzini:2023scu} for different reasons. To further elucidate this point, we shall study the problem of thin shell collapse in PG coordinates using the formalism of Israel junction conditions, which begins with a choice of possibly different coordinate charts on either side of the thin shell. In this context, it will be established that (i) the PG time coordinate is discontinuous across the thin shell, and (ii) this fact is precisely what ensures that the signature of an Israel shock wave, unlike a weak shock wave, is preserved during evolution. 
We conclude by giving a tentative and qualitative picture for the post-bounce effective dynamics in which the shock is treated through the Israel approach.


\section{Weak solutions in dust collapse}
To fix notation, we recall the expression for the metric in spherical symmetry in PG coordinates,
\begin{equation}
    \dd s^2 = -\dd t^2 + \frac{1}{1+E(t,r)}(\dd r + N^r(t,r)\dd t)^2 + r^2d\Omega^2~,
\end{equation}
where $\dd\Omega^2 = \sin^2\theta\dd\theta + \dd\phi^2$, $N^r$ is the radial component of the shift vector and $E$ is related with the total energy of the fluid layers. The equations for gravity coupled to a pressureless fluid (i.e. dust) are given by the general form 
\begin{align}
    \partial_t M + F(M,E)\partial_rM &= 0~, \\
    \partial_t E + F(M,E)\partial_rE &= 0~,
\end{align}
where $M$ is the Arnowitt-Deser-Misner (ADM) mass function and is a function of $N^r$ and $E$. The form of $M$ and $f$ depends on whether one is considering Einstein gravity or some effective theory. For the former \cite{Lasky:2006hq, Nolan:2003wp}, 
\begin{equation}
    F(M,E) = -\sqrt{E + \frac{2M}{r}}~, \quad M = \frac{r}{2}[(N^r)^2-E]~,
\end{equation}
whereas for effective theories \cite{Liu:2025fil, Giesel:2023hys, Fazzini:2023ova, Husain:2022gwp}, the expressions are more complicated, and are such as to produce a bounce in $M$ and $E$ during evolution. In the following, we shall focus on the classical equations, and for simplicity, also restrict ourselves to the marginally bound case, i.e. $E = 0$, which implies that there is only one equation to solve, namely
\begin{equation}
    \partial_t M - \sqrt{\frac{2M}{r}}\partial_rM = 0 ~.
\end{equation}
A simple calculation reveals that this equation can be recast as
\begin{equation}
    \partial_t (r^{1/2}M) - \partial_r\left(\frac{(2M)^{3/2}}{3} \right) = 0~. \label{cons-eq-mass}
\end{equation}
Alternatively, one may rewrite in terms of the radial shift $N^r$ in the following way,
\begin{equation}
    \partial_t(rN^r) - \partial_r\left(\frac{r(N^r)^2}{2}\right) = 0~. \label{cons-eq-shift}
\end{equation}
Both of these equations are of the general conservation form,
\begin{equation}
    \partial_t u(t,r) + \partial_r f(r,u(t,r)) = 0~. \label{diff-cons-law}
\end{equation}
The name stems from the fact that in any compact region of space $[r_1, r_2]$,
\begin{equation}
    \frac{\dd}{\dd t}\int_{r_1}^{r_2} dr\, u(t,r) =  f(r_2, u(t, r_2)) - f(r_1, u(t,r_1))~,  \label{integrated-cons-law}
\end{equation}
which means that the quantity $\int_{r_1}^{r_2}dr\, u(t,r)$ is conserved in the sense that it can change due only to the flux $f(r,u)$ through the boundaries $r_1$ and $r_2$. 

Hyperbolic conservation laws have the property that solutions to smooth initial data become discontinuous in a finite time; in dust collapse, when such discontinuities arise in the matter region, they correspond to points where shells of dust have crossed each other \cite{Husain:2022gwp, Fazzini:2023ova}, the so-called (physical) shell-crossing singularities (SCS). Evolution of the system beyond such points can be studied using the integrated equation above. 
More precisely, suppose that there is a shell crossing singularity at $r=R(t)$, and suppose that $\lim_{r\to R^{\pm}}u=u^{\pm}$; then it can be shown \cite{leveque-92} using \eqref{integrated-cons-law} that the discontinuity at $r=R(t)$ propagates according to the equation
\begin{equation}
    \frac{\dd R}{\dd t} = \frac{f^+-f^-}{u^+-u^-} \equiv \frac{[f]^+_{-}}{[u]^+_{-}}~. \label{rh-condition}
\end{equation}
This is known as the Rankine-Hugoniot jump condition and its solution gives the trajectory of the discontinuity in $u$ at $r=R(t)$; this discontinuity is called a shock wave, and is said to describe a weak solution to the conservation law \eqref{diff-cons-law}. It is important to notice here that the weak solution containing the shock is written in a single PG chart, which implies a continuous PG time across the shock. It is clear from this equation that two conservation laws which are related by a change of variables may give rise to distinct shock speeds even though they would be equivalent as \textit{differential equations}. For instance, Eqs. \eqref{cons-eq-mass} and \eqref{cons-eq-shift} are both equivalent as differential equations, and hence admit the same smooth solutions, but the shock speed for each, as determined \eqref{rh-condition} is different. For \eqref{cons-eq-mass}, we find
\begin{equation}
    \frac{\dd R}{\dd t} = -\frac{2}{3}\frac{(N^{r+})^3-(N^{r-})^3}{(N^{r+})^2-(N^{r-})^2}~, \label{weak-sol-discont}
\end{equation}
whereas \eqref{cons-eq-shift} gives rise to
\begin{equation}
    \frac{\dd R}{\dd t} = -\frac{N^{r+}+N^{r-}}{2}~. \label{weak-sol-cont}
\end{equation}
This shows that given some initial data, a weak solution to a conservation is not unique; there are as many weak solutions for the same initial data as there are ways of writing down the conservation law through a mere change of variables \cite{Nolan:2003wp, leveque-92, Fazzini:2025hsf}. 

While there is no a priori way of preferring one solution over another, in the case of marginally bound dust collapse, \eqref{weak-sol-cont} has a crucial advantage over others, namely that the metric is continuous across the shock at $r=R(t)$ if and only if $\dd R/\dd t$ is given by \eqref{weak-sol-cont} \cite{Husain:2025wrh, Liu:2025fil}. This can be confirmed by a short calculation. Since $dr|_{r=R} = \frac{\dd R}{\dd t}\dd t$, the metric induced on the shock from either side reads 
\begin{align}
    \dd s^2|^+ &= [-1 + (\Dot{R} + N^r{^+})^2] \dd t^2 + R^2\dd \Omega^2~, \label{first}\\
    \dd s^2|^- &= [-1 + (\Dot{R} + N^{r-})^2] \dd t^2 + R^2\dd \Omega^2~. \label{second}
\end{align}
Continuity of the metric requires $\dd s^2|^+=\dd s^2|^-$, which holds if and only if \eqref{weak-sol-cont} does, as can be verified by direct substitution. 

One might think this argument as yielding a decisive way of picking out a unique solution to the problem of the dynamical extension of spacetime beyond SCS out of an infinite number of weak solutions. However, the proposed remedy has a problem. To bring it into relief, consider an initial collapsing thin shell of dust. That is, the geometry inside the shell is Minkowski, so that $N^{r-}=0$, whereas the exterior geometry is Schwarzschild, so that $N^{r+}=\sqrt{R_S/r}$, where $R_S = 2m$ is the Schwarzschild radius, and $M$ the mass of the shell. Then \eqref{weak-sol-cont} reduces to 
\begin{equation}
\frac{\dd R(t)}{\dd t}=-\frac{(N^{r+}+N^{r-})}{2}=-\sqrt{\frac{R_S}{4R}}, \label{vel}
\end{equation}
Introducing coordinates $(\tau,\theta,\phi)$ on the falling shell, where $\tau$ is the proper time of an observer comoving with the shell, the three-metric on the shell can be written as
\begin{equation}
    \dd s^2 = -\dd\tau^2 + R(\tau)^2\dd\Omega^2~.
\end{equation}
This allows us to re-express \eqref{vel} in terms of the proper time of the shell: 
\begin{equation}
\left(\frac{\dd R(\tau)}{\dd \tau} \right)^2=\frac{1}{\frac{4R(\tau)}{R_S}-1} ~.  
\label{Hassandyn}
\end{equation}
Notice that this expression becomes divergent for $R(\tau)=\frac{R_S}{4}$. Since $\tau$ is the shell proper time, and not a coordinate time, it signals that some of the assumptions used to derive the equation might be wrong. Investigating this is the task of the next section.

\section{Signature change and metric continuity in PG coordinates}
A different way of looking at the problem identified above is to notice that if we substitute the shock velocity $\Dot{R}$ (computed with respect to the PG time variables $t$), in, for example, \eqref{first}, we obtain
\begin{equation}
 \dd s^2=\left(-1+ \frac{R_S}{4R}\right)\dd t^2 + R^2 \dd\Omega^2 ~.
 \label{wrong}
\end{equation}
It is immediate to notice that the induced metric as seen from both the interior and exterior changes signature during the evolution of the shell, and since $\dd s^2$ gives the proper time interval of the shell (when $\theta $ and $\phi $ are fixed), this leads to a proper time interval that becomes null when $R=R_S/4$, and changes sign for $R<R_S/4$. As pointed out in \cite{Mars:1993mj, Senovilla:2018hrw} hypersurfaces that change signatures can arise in solutions of Einstein theory (or its distributional form), but whether they provide nonphysical consequences or not has to be checked case by case. In our case, this implies that when the shell is at $R>R_S/4$ ($R<R_S/4$), then only a time-like (space-like) particle can co-move together with the shell surface, while when $R=R_S/4$ only a null particle can. Even if this kind of spacetime can be mathematically defined, it seems unrealistic from a physical perspective, since the shock is generated by a thin shell of dust which should move in a time-like fashion.

Since the only requirement to find the shock velocity \eqref{vel} is continuity of the induced metric in PG coordinates --- a plausible assumption at first sight---the question is how the dynamics can be nonphysical. 
The reason is subtle and concerns the behavior of the PG time coordinate across the shock. The discussion so far is based on the assumption that a single PG chart can be used both for the spacetime interior to the thin shell (which trivially gives the Minkowski metric), and for the exterior (that gives the Schwarzschild metric in PG coordinates). However, the time flowing in these two spacetimes, say $t_+$ for the exterior and $t_{-}$ for the interior, is extremely different; this is evident in the fact that a Schwarzschild metric in PG coordinates causes gravitational time dilation, whereas a Minkowski metric does not. When we require equality between \eqref{first} and \eqref{second}, we require not only continuity of the induced metric on the shell, but also that $\dd t_-^2= \dd t_+^2$, which means that the flow of PG time is the same as measured from both sides of the shell. This is a stronger assumption than mere metric continuity, since it does not account for a possible coordinate discontinuity of the PG time at the shock, and forces the signature of the shock induced metric to change during the evolution. 


\section{PG time and Israel junction conditions}
In light of the foregoing remarks, the only possibility to ensure a signature-preserving dynamics \textit{and} metric continuity at the same time seems to allow the interior and exterior PG times $t_{-}$ and $t_+$ to be different. A natural avenue to explore this point is to study the dynamics of thin shell using Israel junction conditions \cite{Poisson:2009pwt, Israel:1966rt}, and compare the outcome with the previous result. In this formalism, one starts by introducing separate coordinates on either side of the shell. Imposing metric continuity (the first junction condition) allows the Einstein equations to hold in a distributional sense, which gives rise to jump relationships between the extrinsic curvature of the shell and its stress-energy tensor. Together, these conditions then serve to fix the dynamics of the shell completely. If this calculation is done in PG coordinates, assuming the coordinates $\{t_{\pm}, r_\pm,\theta,\phi\}$ on the two sides of the shell, one can see quite explicitly that (i) $t_{+}$ is not the same as $t_{-}$ and (ii) the shell always remains timelike. Since the latter fact gives us good reason to believe that the dynamics is physical, we are apt to take the equality of $t_{+}$ and $t_{-}$, which has to be assumed to get a weak solution with continuous induced metric, with a grain of salt. Thus, it will be instructive to run through the formalism of Israel junction conditions applied to a thin shell.

To begin with, instead of \eqref{first} and \eqref{second}, one starts by writing
\begin{align}
    \dd s^2|^+ &= \left[-1 + \left(\frac{\dd R_+}{\dd t_+} + N^r{^+}\right)^2\right] \dd t_+^{2} + R_+^2\dd \Omega^2~,\\
    \dd s^2|^- &= \left[-1 + \left(\frac{\dd R_-}{\dd t_-} + N^{r-}\right)^2\right] \dd t_-^2 + R_-^2\dd \Omega^2~.
\end{align}

By equating the two line elements, one gets $R_+=R_-\equiv R$, and
\begin{equation}
\left( \frac{\dd t_+}{\dd t_-}\right)^2 =\frac{-1+\left(N^{r-}+\frac{\dd R}{\dd t_-}\right)^2}{-1+\left(N^{r+}+\frac{\dd R}{\dd t_+}\right)^2}  ~.
\end{equation}

Moreover, since at $\theta,\phi$ fixed and on the shell location, the line element should give the shell proper time flowing, we require $\dd s^2 =-\dd \tau^2$. It follows that
\begin{align}
 &-1=\left[-1+\left(\frac{\dd R}{\dd t_+}+ N^r{^+}\right)^2\right]\frac{\dd t_+^2}{\dd \tau^2} ~, \label{first1}\\
 &-1=\left[-1+\left(\frac{\dd R}{\dd t_-} + N^r{^-}\right)^2\right]\frac{\dd t_-^2}{\dd \tau^2} ~. \label{first2}
 \end{align}

Notice that by requiring $\dd s^2|_{\theta,\phi=const.}=-\dd \tau^2$ implies signature preservation of the induced metric at the shell.
In other words, for metric continuity to be compatible with a signature-preserving metric, the PG times for the exterior and interior must be related to the shock velocities with respect to the exterior and interior by the above equations. However, this equation now does not fix the shock velocity. To fix the dynamics completely, one has to use the relation between the induced stress tensor and the extrinsic curvature of the shell to fix the dynamics. Since the expression for the extrinsic curvature of the shell is quite complicated in PG coordinates, we will instead write the interior and exterior metric in Minkowski and Schwarzschild coordinates respectively, and after having obtained an equation for the shell velocity, revert to PG coordinates to exhibit the relation between the interior and exterior PG times. Thus, we write 
\begin{align}
    & \dd s^2=-\dd T_-^2+\dd r_-^2 +r_-^2\dd \Omega^2 ~,\\
    & \dd s_+^2=-F(r_+) \dd T_+^2 +\frac{1}{F(r_+)}\dd r_+^2+r_+^2 \dd \Omega^2 ~,
\end{align}
with $F\equiv 1-{R_S}/{r_+}$. Notice that here $t_+$ is the Schwarzschild time coordinate. 
For the induced metric on the shell hypersurface, as before, we take as coordinates $\{\tau, \theta, \phi \}$, where $\tau$ is the proper time of the shell. The shell dynamics, as seen from the interior will be given by $\{T_-=T_-(\tau), r_-=R_-(\tau)\}$, while as seen from the exterior, by $\{T_+=T_+(\tau), r_-=R_+(\tau)\}$.
Now, the induced metric on the shell, as computed using the interior metric, is given by
\begin{equation}
  \dd s^{+2}_\Sigma=(-\Dot{T}_-^2+\Dot{R}_-^2)\dd \tau^2 +R_-^2 \dd \Omega^2 ~,
\end{equation}
while, as computed from the exterior, it is
\begin{equation}
  \dd s^{+2}_\Sigma=\left(- \Dot{T}_-^2 ~F(R_+(\tau))+\Dot{R}_-^2\frac{1}{F(R_+(\tau))}\right)\dd \tau^2 +R_+^2 \dd \Omega^2~.
\end{equation}
Since $\tau$ is the shell proper time, we need the following to be satisfied
\begin{equation}
    \begin{cases}
    &-\Dot{T}_-^2+\Dot{R}_-^2 =-1\\
    &- F(R_+(\tau))\Dot{T}_+^2 +\Dot{R}_+^2\frac{1}{F(R_+(\tau))}  =-1      
    \end{cases}
    \label{1junction}
\end{equation}
Moreover, to get continuity of the induced metric on the shell $\Sigma$, we need $R_+(\tau)=R_-(\tau)\equiv R(\tau)$. Notice that, as mentioned before, requiring \eqref{1junction} forces the induced metric at the shell not to change signature along the dynamics, in contrast to what happens in \eqref{wrong}. 
Supplemented by the relation between the jump in the extrinsic curvature and the surface stress-energy tensor \cite{Poisson:2009pwt} (i.e. a violation of the second Israel junction condition), Eq. \eqref{1junction} provides a unique dynamical equation for the shell radius, given by \cite{Poisson:2009pwt}
\begin{equation}
\Dot{R}^2=\frac{R_S}{2R} \left(\frac{R_S}{8R}+1 \right) ~,
\label{eqpoisson}
\end{equation}
for a thin shell starting at $R=+\infty$ with zero initial kinetic energy (marginally bound case). Notice that this equation is well defined for any $R$, and describes a collapsing dynamics ending in the Schwarzschild central singularity. From \eqref{1junction}, one can easily get
\begin{align}
    & {\Dot{T}_+} = \frac{\sqrt{F+\Dot{R}^2}}{F} ~,   \\
    &\Dot{T}_-=\sqrt{1+\Dot{R}^2}~.
    \label{tminus}
\end{align}
The first equation tells us how the exterior Schwarzschild time $T_+$ flows during the shell evolution, while the second describes how the interior Minkowski time $T_-$ flows. 
Since Minkowski time is the same as PG time for the interior flat geometry, in order to compare the PG times in the exterior and interior, one can simply switch to PG time in the exterior. To this end, recall that for the advanced PG time coordinate, we have
\begin{equation}
t^{PG}_+=t_++\int \sqrt{\frac{R_S}{r^+}}\frac{1}{F(r^+)}\dd r^+~. 
\end{equation}
Thus we get
\begin{align}
 \Dot{T}_+^{PG}=&\Dot{T}_+\frac{\dd t^{PG}_+}{\dd t_+}\bigg|_{r^+=R(\tau)}      = \Dot{T}_+ \left(1+ \frac{1}{1-\frac{R_S}{R}}\sqrt{\frac{R_S}{R}}\frac{\dd R}{\dd t_+} \right)~.    
\end{align}
We can rewrite the last term in terms of $\Dot{R}$, through
\begin{equation}
 \frac{\dd R}{\dd t_+}= \frac{\Dot{R}}{\Dot{T}_+}  ~,
\end{equation}
obtaining
\begin{align}
\Dot{T}^{PG}_+=&\Dot{T}_+ + \frac{\Dot{R}}{1-\frac{R_S}{R}}\sqrt{\frac{R_S}{R}}= \notag \\
             =&\frac{\sqrt{1-\frac{R_S}{R}+\Dot{R}^2}}{1-\frac{R_S}{R}}+\frac{\Dot{R}}{1-\frac{R_S}{R}}\sqrt{\frac{R_S}{R}} ~.
             \label{tpg}
\end{align}
Finally, upon comparing \eqref{tminus} with \eqref{tpg}, and imposing equality between them, we find
\begin{equation}
    \left( 1-\frac{R_S}{R}\right)\sqrt{1+\Dot{R}^2}={\sqrt{1-\frac{R_S}{R}+\Dot{R}^2}}+{\Dot{R}}\sqrt{\frac{R_S}{R}}~.
    \end{equation}
One can check that this equation is not satisfied for $\Dot{R}$ given by \eqref{eqpoisson}. This implies that the PG time is discontinuous across the Israel thin shell, and thus one cannot use a single PG chart to study the shell's dynamics, as we advertised earlier. Since equations \eqref{1junction} implement  continuity in the induced metric on the shell without allowing signature change, and the dynamics provided by \eqref{eqpoisson} comes from the correct relation between the surface energy-momentum tensor and the extrinsic curvature on the shell (see \cite{Poisson:2009pwt} for the detailed computation), then we can safely conclude that the Israel solution provides the only possibility to get continuity of the induced metric and preserve the metric signature, accounting correctly for the discontinuity in the PG time across the shock. 

\section{Implications for weak solutions}
The result obtained in the previous section shows that the construction of a weak solution for the thin shell collapse by requiring continuity as done in \cite{Husain:2025wrh}, leads to a result different from the one of Israel, with a shock velocity given by \eqref{vel}. As argued in the previous two sections, the dynamics from the weak solution should be considered wrong from a physical perspective. This conclusion could be extended beyond the classical thin shell collapse. If the weak approach does not work for the simplest shock dynamics---a shock in the initial data---it cannot be relied upon for more complicated dynamics.
One can start with, say, a Gaussian packet, which develops SCS during its dynamics. The Rankine-Hugoniot condition \eqref{rh-condition} tells us that the SCS is replaced by a propagating discontinuity. Since the weak solution then follows a dynamics that does not allow the PG time to be discontinuous and does not enforce signature preservation for the induced metric, the conclusions reached above apply. Notice, indeed, that even for initial profiles that are not isolated thin shells, when SCS form the mass function becomes discontinuous, and the physical considerations concerning the discontinuity of PG time hold.


The above arguments raise important concerns about weak solutions in general relativistic dust collapse \cite{Nolan:2003wp,Lasky:2006hq}, and in effective theories like the one constructed in \cite{Husain:2022gwp,Giesel:2023hys} that admit SCS formation. As was pointed out above, weak solutions are not unique; infinitely many weak solutions can be constructed out of the same original differential equation that governs the system. A natural choice to choose between weak solutions, as pointed out in \cite{Husain:2025wrh} in the classical case, and \cite{Liu:2025fil} in the effective case, is to require continuity of the induced metric when shell-crossing singularities form. However, as proved in the previous sections, this gives rise to nonphysical dynamics, and hence the shock velocity provided by the Rankine-Hugoniot condition is not reasonable from a physical point of view. This conclusion, even if it can be proved only for the weak solution with a continuous induced metric at the shock, holds conceptually also for other, discontinuous, weak solutions. 
The general argument above concerning the discontinuity of PG time across the shock would hold regardless of discontinuity in the induced metric, so the particular integral form of the effective PDE considered. 
 
Finally, notice that the above considerations apply not only to effective LQG-inspired dynamics that develop shell-crossing singularities before the bounce, but also the ones that develop shocks after the bounce \cite{Husain:2021ojz,Husain:2022gwp,Giesel:2023hys,Cipriani:2024nhx,Fazzini:2023ova,Fazzini:2025hsf,Liu:2025fil,Bobula:2024chr}--although the initial data are different, the way SCS are treated is the same.
By the same token, if one recognizes that the signature changing is a signal of an nonphysical dynamics, then one also cannot trust the dynamics given by \eqref{vel} for initial data that do not give rise to a signature-changing shock---since the shock is treated in the same way--- as for collapsing stars which have a very small mass at the bounce \cite{Liu:2025fil}. 

\section{A different post-bounce dynamics for the shock wave model}

Based on the results derived in the previous sections, it turns out that the most physical shock dynamics is the one given by the Israel junction conditions, both at the classical and effective level. If we consider therefore stellar collapse in effective LQG that develops shell-crossing singularities after the bounce \cite{Fazzini:2023ova, Fazzini:2025hsf}, then the shock velocity should be given by the first Israel junction conditions, supplemented with a modified (effective) relation between the jump in the extrinsic curvature and the energy-momentum tensor of the (non isolated) thin shell \cite{Giesel:2023hys}, that takes account of quantum modifications of Einstein equations. Conventionally, the formalism of Israel junction conditions is useful when one knows the spacetime geometry on either side of a shock. For instance, in the case of classical thin-shell collapse considered above, one knows the interior and exterior to be Minkowski and Schwarzschild, respectively. In order to extend this formalism to the general case where the geometry on either side of the shock is not given a priori but has to be determined by solving the equations of motion, one would presumably need to implement the junction conditions numerically at every instance of a shell crossing, the solution on either side of the crossing being determined by numerically integrating the equations of motion. We leave a detailed study along these lines for future work.

Even if the exact evolution needs numerics, since the energy density profile at SCS is not the one of an isolated thin shell, we can give a qualitative picture of this dynamics. Immediately after the shock forms, since it is forced to move time-like (this is guaranteed by the first Israel junction condition for time-like shells), it cannot move toward the trapped region, as in \cite{Husain:2022gwp}. In the meanwhile, stellar matter inside and outside the shock will accumulate on the shock, since the matter inside moves outward, and the one outside the shock moves inward. Therefore, soon after the bounce, the shock will acquire the whole mass of the original star, becoming an isolated thin shell. 

The exterior vacuum, differently from the classical Schwarzschild vacuum, will bounce \cite{Lewandowski:2022zce,Bobula:2024chr}, producing firstly a non-trapped region, and then an anti-trapped region (notice that the anti-trapped region cannot be recovered through a single advanced PG chart, but requires a maximal extension, similar to the classical Reissner-Nordström black hole). When the vacuum develops a non-trapped region and then the anti-trapped region, the shock will move outward.
The dynamics of the shock, soon after the bounce, can be approximated by the Israel time-reversed dynamics of the classical thin shell-collapse---since it soon leaves the deep quantum region---with the proper $3$-velocity given by \eqref{eqpoisson}. This means that the shock will reach the white hole outer horizon, eventually moving toward a second asymptotic region. This picture, which at this stage has to be considered heuristic, will be studied rigorously in a future work.

Another approach to this problem, developed in the literature, deserves attention. It has been shown \cite{Friedman:1997gr,Fiamberti:2007} that one can recover the classical Israel dynamics for the thin shell starting from a variational principle, using an action that includes boundary terms properly accounting for the thin-shell matter content. This approach has recently been employed to derive an effective thin-shell dynamics \cite{Sahlmann:2025fde}. While the resulting dynamics reproduces the correct classical limit, it cannot be considered fully reliable, as it suffers from issues similar to those of the weak approach. In particular, it does not allow the imposition of the first Israel junction conditions \eqref{first1}, \eqref{first2} in the deep quantum region, which in turn leads to a complex-valued lapse function. Even when a relaxed version of these conditions is imposed, the time-like behavior of the thin shell---which is guaranteed automatically in the Israel framework---is recovered only for small black hole masses. Future studies will aim to clarify why this approach works at the classical level but fails at the effective level.

\section{Conclusion}
When one deals with a non-linear PDE in conservation form, the natural way of handling them is to look at weak solutions. This approach allows us to study the dynamics when the fields develop discontinuities, a typical behavior for solutions of these kinds of equation. In physics, a large class of phenomena is described extremely well by weak solutions, such as sonic booms, Bose-Einstein condensate dynamics, supernova explosions, and so on \cite{Dafermos:1315649}. It is thus natural to ask: why does the approach not work for gravitational dust collapse? We have seen that during gravitational collapse studied in PG coordinates, when the solution develops a shock, and the induced metric at the shock is continuous (so that an unambiguous signature can be assigned to the shock surface), the shock surface changes signature during evolution. This fact, as we have seen, arises from the underlying assumption that the PG time flows in the same manner on either side of the shock. If one relaxes this assumption, at least in the case of a thin-shell shock wave, one can obtain a more realistic evolution that preserves the signature by using junction conditions. These considerations support the conclusion that it is ill-advised to use a single PG coordinate chart to cover the whole of a spacetime containing SCS. Since weak solutions of dust collapse cannot do otherwise, it seems that one needs to go beyond them in order to have a more physical picture of SCS as propagating shock waves. This issue does not arise in the other applications of weak solutions mentioned above because the time coordinate there is well-defined everywhere and does not develop discontinuities, so that using a single coordinate patch for the whole dynamics is always meaningful. 

A possible way to overcome this issue and study the dynamics beyond shell-crossing singularities in Einstein theory or in effective theories is to construct an integral equation that accounts for this discontinuity in the time coordinate, possibly leading to a relativistic version of the Rankine-Hugoniot jump condition. Another possibility would be to modify the underlying theory (e.g. with terms that account for other quantum gravitational effects than the ones studied in \cite{Husain:2022gwp,Giesel:2023hys,Cipriani:2024nhx,Fazzini:2023ova,Fazzini:2025hsf,Liu:2025fil}, that avoid the shell-crossing singularity formation and thereby avoid the need of introducing weak solutions at all. Yet another approach might be to generalize the formalism of Israel junction conditions to study arbitrary distributions of matter. This would presumably involve implementing the junction conditions in a numerical code, where at each time step the shock velocity and the induced metric are computed using the values of the fields on either side of the shock. Preliminary tentative considerations show that the effective shock dynamics soon after the bounce can be approximated as a time-reversed thin shell collapse. The shock will exit the white hole outer horizon and reach the second asymptotic region.
A detailed investigation of this model is left for future work.


\acknowledgments
The authors thank Viqar Husain, Edward Wilson-Ewing and Michał Bobula for helpful comments.
This work is supported in part by the Natural Sciences and Engineering Research Council of Canada.


\begin{thebibliography}{10}

\bibitem{Oppenheimer:1939ue}
J. R. Oppenheimer and H. Snyder, \textit{On Continued gravitational contraction}, \textit{Phys. Rev.} \textbf{56}: 455--459 (1939), \href{https://doi.org/10.1103/PhysRev.56.455}{DOI}.

\bibitem{Tolman:1934za}
R. C. Tolman, \textit{Effect of imhomogeneity on cosmological models}, \textit{Proc. Nat. Acad. Sci.} \textbf{20}: 169--176 (1934), \href{https://doi.org/10.1073/pnas.20.3.169}{DOI}.

\bibitem{Bondi:1947fta}
H. Bondi, \textit{Spherically symmetrical models in general relativity}, \textit{Mon. Not. Roy. Astron. Soc.} \textbf{107}: 410--425 (1947), \href{https://doi.org/10.1093/mnras/107.5-6.410}{DOI}.

\bibitem{Husain:2022gwp}
V. Husain, J. G. Kelly, R. Santacruz and E. Wilson-Ewing, \textit{Fate of quantum black holes}, \textit{Phys. Rev. D} \textbf{106}(2): 024014 (2022), \href{https://doi.org/10.1103/PhysRevD.106.024014}{DOI}, arXiv:2203.04238 [gr-qc].

\bibitem{Bobula:2023kbo}
M. Bobula and T. Pawłowski, \textit{Rainbow Oppenheimer-Snyder collapse and the entanglement entropy production}, Phys. Rev. D \textbf{108} (2), 026016 (2023), arXiv:2303.12708, \href{https://doi.org/10.1103/PhysRevD.108.026016}{DOI}.

\bibitem{Lewandowski:2022zce}
J. Lewandowski, Y. Ma, J. Yang, and C. Zhang, \textit{Quantum Oppenheimer-Snyder and Swiss Cheese Models}, Phys. Rev. Lett. \textbf{130} (10), 101501 (2023), arXiv:2210.02253, \href{https://doi.org/10.1103/PhysRevLett.130.101501}{DOI}.


\bibitem{Giesel:2023hys}
K. Giesel, H. Liu, P. Singh and S. A. Weigl, \textit{Generalized analysis of a dust collapse in effective loop quantum gravity: Fate of shocks and covariance}, \textit{Phys. Rev. D} \textbf{110}(10): 104016 (2024), \href{https://doi.org/10.1103/PhysRevD.110.104016}{DOI}, arXiv:2308.10953 [gr-qc].

\bibitem{Cipriani:2024nhx}
L. Cipriani, F. Fazzini and E. Wilson-Ewing, \textit{Gravitational collapse in effective loop quantum gravity: Beyond marginally bound configurations}, \textit{Phys. Rev. D} \textbf{110}(6): 066004 (2024), \href{https://doi.org/10.1103/PhysRevD.110.066004}{DOI}, arXiv:2404.04192 [gr-qc].


\bibitem{Husain:2021ojz}
V. Husain, J. G. Kelly, R. Santacruz and E. Wilson-Ewing, \textit{Quantum Gravity of Dust Collapse: Shock Waves from Black Holes}, \textit{Phys. Rev. Lett.} \textbf{128}(12): 121301 (2022), \href{https://doi.org/10.1103/PhysRevLett.128.121301}{DOI}, arXiv:2109.08667 [gr-qc].

\bibitem{Alonso-Bardaji:2023qgu}
A. Alonso-Bardaji and D. Brizuela, \textit{Nonsingular collapse of a spherical dust cloud}, Phys. Rev. D \textbf{109} (6), 064023 (2024), arXiv:2312.15505, \href{https://doi.org/10.1103/PhysRevD.109.064023}{DOI}.

\bibitem{Han:2023wxg}
M. Han, C. Rovelli, and F. Soltani, \textit{Geometry of the black-to-white hole transition within a single asymptotic region}, Phys. Rev. D \textbf{107} (6), 064011 (2023), arXiv:2302.03872, \href{https://doi.org/10.1103/PhysRevD.107.064011}{DOI}.

\bibitem{Cafaro:2024vrw}
L. Cafaro and J. Lewandowski, \textit{Status of Birkhoff's theorem in the polymerized semiclassical regime of loop quantum gravity}, Phys. Rev. D \textbf{110} (2), 024072 (2024), arXiv:2403.01910, \href{https://doi.org/10.1103/PhysRevD.110.024072}{DOI}.


\bibitem{hu}
H. Hu, \textit{Exact solutions of the spherically symmetric gravitational field equations}, \textit{Front. Math. China} \textbf{1}: 169–177 (2006).

\bibitem{chao-hao}
G. Chao-hao, \textit{Gravitation collapse of spherical symmetry with non-uniform density}, \textit{Front. Math. China} \textbf{1}: 161--168 (2006).

\bibitem{yodzis1973occurrence}
P. Yodzis, H. J. Seifert and H. Müller Zum Hagen, \textit{On the occurrence of naked singularities in general relativity}, \textit{Commun. Math. Phys.} \textbf{34}(2): 135--148 (1973).

\bibitem{Nolan:2003wp}
B. C. Nolan, \textit{Dynamical extensions for shell crossing singularities}, \textit{Class. Quant. Grav.} \textbf{20}: 575--586 (2003), \href{https://doi.org/10.1088/0264-9381/20/4/302}{DOI}, arXiv:gr-qc/0301028.

\bibitem{Lasky:2006hq}
P. D. Lasky, A. W. C. Lun and R. B. Burston, \textit{Initial value formalism for dust collapse}, arXiv:gr-qc/0606003 (2006).


\bibitem{Martel:2000rn}
K. Martel and E. Poisson, \textit{Regular coordinate systems for Schwarzschild and other spherical space-times}, \textit{Am. J. Phys.} \textbf{69}: 476--480 (2001), \href{https://doi.org/10.1119/1.1336836}{DOI}, arXiv:gr-qc/0001069.

\bibitem{leveque-92}
R. J. LeVeque, \textit{Numerical methods for conservation laws (2. ed.)}, Birkhäuser, Lectures in mathematics, 1992, pp. 1--214, ISBN: 978-3-7643-2723-1.

\bibitem{Dafermos:1315649}
C. M. Dafermos, \textit{Hyperbolic Conservation Laws in Continuum Physics; 3rd ed.}, Springer, Dordrecht, Grundlehren der mathematischen Wissenschaften, 2010, \href{https://doi.org/10.1007/978-3-642-04048-1}{DOI}, \url{https://cds.cern.ch/record/1315649}.


\bibitem{Fazzini:2023ova}
F. Fazzini, V. Husain and E. Wilson-Ewing, \textit{Shell-crossings and shock formation during gravitational collapse in effective loop quantum gravity}, \textit{Phys. Rev. D} \textbf{109}(8): 084052 (2024), \href{https://doi.org/10.1103/PhysRevD.109.084052}{DOI}, arXiv:2312.02032 [gr-qc].


\bibitem{colombeau-92}
J. F. Colombeau, \textit{Multiplication of distributions. A tool in mathematics, numerical engineering and theoretical physics}, Lecture Notes in Mathematics, vol. 1532, Springer-Verlag, Berlin, 1992, ISBN: 3-540-56288-5, \href{https://doi.org/10.1007/BFb0088952}{DOI}.


\bibitem{Fazzini:2025hsf}
F. Fazzini, \textit{Non-uniqueness of the shockwave dynamics in effective loop quantum gravity}, arXiv:2502.03003 [gr-qc] (2025).

\bibitem{Husain:2025wrh}
V. Husain and H. Mehmood, \textit{Shock waves in classical dust collapse}, arXiv:2504.14883 [gr-qc] (2025).

\bibitem{Liu:2025fil}
H. Liu and D. Qu, \textit{Quantum induced shock dynamics in gravitational collapse: insights from effective models and numerical frameworks}, arXiv:2504.18462 [gr-qc] (2025).

\bibitem{Fazzini:2023scu}
F. Fazzini, C. Rovelli and F. Soltani, \textit{Painlevé-Gullstrand coordinates discontinuity in the quantum Oppenheimer-Snyder model}, \textit{Phys. Rev. D} \textbf{108}(4): 044009 (2023), \href{https://doi.org/10.1103/PhysRevD.108.044009}{DOI}.

\bibitem{Mars:1993mj}
M. Mars and J. M. M. Senovilla, \textit{Geometry of general hypersurfaces in space-time: Junction conditions}, \textit{Class. Quant. Grav.} \textbf{10}: 1865--1897 (1993), \href{https://doi.org/10.1088/0264-9381/10/9/026}{DOI}, arXiv:gr-qc/0201054.


\bibitem{Senovilla:2018hrw}
J. M. M. Senovilla, \textit{Equations for general shells}, \textit{JHEP} \textbf{11}: 134 (2018), \href{https://doi.org/10.1007/JHEP11(2018)134}{DOI}, arXiv:1805.03582 [gr-qc].

\bibitem{Poisson:2009pwt}
E. Poisson, \textit{A Relativist's Toolkit: The Mathematics of Black-Hole Mechanics}, Cambridge University Press (2009), \href{https://doi.org/10.1017/CBO9780511606601}{DOI}.

\bibitem{Israel:1966rt}
W. Israel, \textit{Singular hypersurfaces and thin shells in general relativity}, Nuovo Cim. B \textbf{44S10}, 1 (1966) [Erratum: Nuovo Cim.B 48, 463 (1967)], \href{https://doi.org/10.1007/BF02710419}{DOI}.



\bibitem{Israel:1967zz}
W. Israel, \textit{Gravitational Collapse and Causality}, Phys. Rev. \textbf{153}, 1388--1393 (1967), \href{https://doi.org/10.1103/PhysRev.153.1388}{DOI}.



\bibitem{Bobula:2024chr}
M. Bobula and T. Pawłowski, \textit{Causal structure of nonhomogeneous dust collapse in effective loop quantum gravity}, arXiv:2410.22943 (2024).

\bibitem{Friedman:1997gr}
J.~L.~Friedman, J.~Louko, and S.~N.~Winters-Hilt,
``Reduced phase space formalism for spherically symmetric geometry with a massive dust shell,''
Phys.\ Rev.\ D {\bf 56}, 7674--7691 (1997).

\bibitem{Fiamberti:2007}
F.~Fiamberti and P.~Menotti,
``Reduced Hamiltonian for intersecting shells,''
Nucl.\ Phys.\ B {\bf 794}, 512--537 (2008).


\bibitem{Sahlmann:2025fde}
H.~Sahlmann and C.~Zhang,
``Dust shell in effective loop quantum black hole model,''
arXiv:2506.04589 [gr-qc], Jun 2025.


\end{thebibliography}
\end{document}